\documentclass{elsart}
\usepackage{epsfig,amsfonts,here}

\def\Journal#1#2#3#4{{#1} {\bf #2}, #3 (#4)}

\def\SNP{{\em Sov. J. Nucl. Phys.}}
\def\NPB{{\em Nucl. Phys.} B}
\def\PLB{{\em Phys. Lett.}  B}
\def\PRL{\em Phys. Rev. Lett.}
\def\PRD{{\em Phys. Rev.} D}
\def\ZPC{{\em Z. Phys.} C}
\def\EPG{{\em Eur.Phys.J.} C}
\def\EPJL{{\em Europhys. Lett.}}
\def\IJMP{\em Int. Jour. of Mod. Phys.}

\def\be{\begin{equation}}
\def\ee{\end{equation}}
\def\bea{\begin{eqnarray}}
\def\eea{\end{eqnarray}}

\begin{document}
\begin{frontmatter}

\title{SOFT PHYSICS -- THREE POMERONS?}
\author[russia]{V. A. Petrov}, 
\ead{petrov@mx.ihep.su}
\author[italy]{A. V. Prokudin}
\ead{prokudin@to.infn.it}

\address[russia]{
Institute For High Energy Physics,
142281 Protvino,  RUSSIA}
\address[italy]{
Dipartimento di Fisica Teorica, 
Universit\`a Degli Studi Di Torino, 
Via Pietro Giuria 1,
10125 Torino, 
ITALY
and
Sezione INFN di Torino,
 ITALY}

\begin{abstract}
A model of a three Pomeron contribution to high energy 
elastic $pp$ and $\bar p p$ scattering is proposed. The data are well
described for all momenta ($0.01\le |t|\le 14.\; GeV^2$) and energies 
($8.\le\sqrt{s}\le 1800.\; GeV$) ($\chi^2/{\rm d.o.f.}=2.60$). The model 
predicts the appearance of two dips in the differential cross-section 
which will be measured at LHC. The parameters of the Pomeron trajectories are:
 \\
$\alpha(0)_{{\Bbb P}_1}=1.058,\;\;\alpha'(0)_{{\Bbb P}_1}=0.560\;(GeV^{-2});$ \\
$\alpha(0)_{{\Bbb P}_2}=1.167,\;\;\alpha'(0)_{{\Bbb P}_2}=0.273\;(GeV^{-2});$ \\
$\alpha(0)_{{\Bbb P}_3}=1.203,\;\;\alpha'(0)_{{\Bbb P}_3}=0.094\;(GeV^{-2}).$ 
\end{abstract}
\end{frontmatter}

\section{INTRODUCTION}

Fervently awaited high-energy collisions at LHC will give an access 
not only to yet unexplored
small distances but also simultaneously to neither explored 
large distances~\cite{Petrov}.
Future measurements of total and elastic cross-sections 
at LHC~\cite{multipomeronFaus-Golfe}
tightly related to the latter domain 
naturally stimulate further searches
for new approaches to diffractive scattering at high energies.
 
Recently some models with multi-Pomeron structures were proposed
\cite{{multipomeronNicolescu},{multipomeronKontros},{multipomeronLandshoff}}.
 Some of them \cite{multipomeronNicolescu}, \cite{multipomeronKontros},
 \cite{multipomeronLandshoff} use Born amplitudes with two Pomerons as single 
\cite{multipomeronNicolescu}, \cite{multipomeronLandshoff}, or double poles \cite{multipomeronKontros}.

The eikonal models that are capable of describing the data for nonzero	transferred momenta are developed in Refs \cite{multipomeronPredazzi},
 \cite{multipomeronPancheri}. It is worth noticing that the two-Pomeron eikonal has been applied 
to the description of the data more than ten years ago 
(see, e.g., Ref. \cite{multipomeronLikhoded}).

The very multiformity of the models hints that maybe the most 
general way to describe high-energy diffraction is just to admit 
an arbitrary number of Pomerons (i.e. all vacuum Regge-poles contributing 
non-negligibly at reasonably high energies. Roughly, they should have
intercepts not lower than $1$).

As it seems not possible to describe the data in the framework of
the eikonal approach with presence of one single pole Pomeron 
contribution~\cite{mypaper}, and the two-Pomeron option does not improve
quality of description drastically (more details are given in~\cite{threepomerons}) 
it is fairly natural
to try the next, three-Pomeron, option for the eikonal.
We will see below that this choice appears rather lucky.

\section{THE MODEL}
Let us brifely outline the basic properties of our model. Unitarity condition:
$$
\Im{\rm m}\; T(s,\vec b) \simeq \vert T(s,\vec b)\vert^2 + \eta (s,\vec b)\; ,
$$
where $T(s,\vec b)$ is the scattering amplitude in the impact
representation,
$\vec b$ is the impact parameter, $\eta (s,\vec b)$ is the
contribution of 
inelastic channels, implies the following eikonal form for the scattering amplitude $T(s,\vec b)$
\begin{equation}
T(s,\vec b)=\frac{e^{2i\delta (s,\vec b)}-1}{2i}\; ,
\label{eq:ampl}
\end{equation}
where $\delta (s,\vec b)$ is the
eikonal function. 

The eikonal function is assumed to have simple poles in the complex $J$-plane 
and the
corresponding Regge trajectories are normally being used in the linear 
approximation
$
\alpha(t) = \alpha(0) + \alpha '(0)t\; .
$

In the present model we assume the following representation for the eikonal function:
\be
\delta_{pp}^{\bar p p}(s,b) = \delta^+_{{\Bbb P}_1}(s,b)+
\delta^+_{{\Bbb P}_2}(s,b)+
\delta^+_{{\Bbb P}_3}(s,b)
\mp \delta^-_{\Bbb
O}(s,b)+\delta^+_{
f}(s,b)\mp \delta^-_{\omega}(s,b),
\label{eq:modeleik}
\ee

here $\delta^+_{{\Bbb P}_{1,2,3}}(s,b)$ are Pomeron 
contributions. `$+$' denotes C even trajectories,
`$-$' denotes  C odd trajectories, $\delta^-_{\Bbb O}(s,b)$ is the 
Odderon contribution ;
$\delta^+_{ f}$, 
$\delta^-_{\omega}(s,b)$ are the contributions of secondary Reggeons, $f$
($C=+1$) and
$\omega$ ($C=-1$). The analytical formulae for the eikonal function are given in \cite{threepomerons}.

The parameters of secondary Reggeon trajectories are fixed according to
the parameters obtained from a fit of the meson spectrum
\cite{gd93}, $ \alpha_f(t) = 0.69+0.84 t$, $\alpha_\omega (t) = 0.47+0.93 t$.
All the other trajectories are taken in linear aproximation $\alpha_i(t) = \alpha_i(0)+\alpha_i'(0)t,\;(i={\Bbb P}_1,{\Bbb P}_2,
{\Bbb P}_3,{\Bbb O})$.

\section{RESULTS}

We fit the adjustable parameters over a set of 982 $pp$ and $\bar p p$ 
data of both forward observables (total cross-sections $\sigma_{tot}$,
and $\rho$ -- ratios of real to imaginary part of the amplitude) in the
range $8.\le\sqrt{s}\le 1800.\; GeV$ and angular distributions 
($\frac{d\sigma}{dt}$) in the ranges $23.\le\sqrt{s}\le 1800.\; GeV$,
$0.01\le |t|\le 14.\; GeV^2$. 

\begin{figure}[H]
\centering
\parbox[t]{6.cm}{\vspace*{ 0cm} \epsfxsize=70mm \epsffile{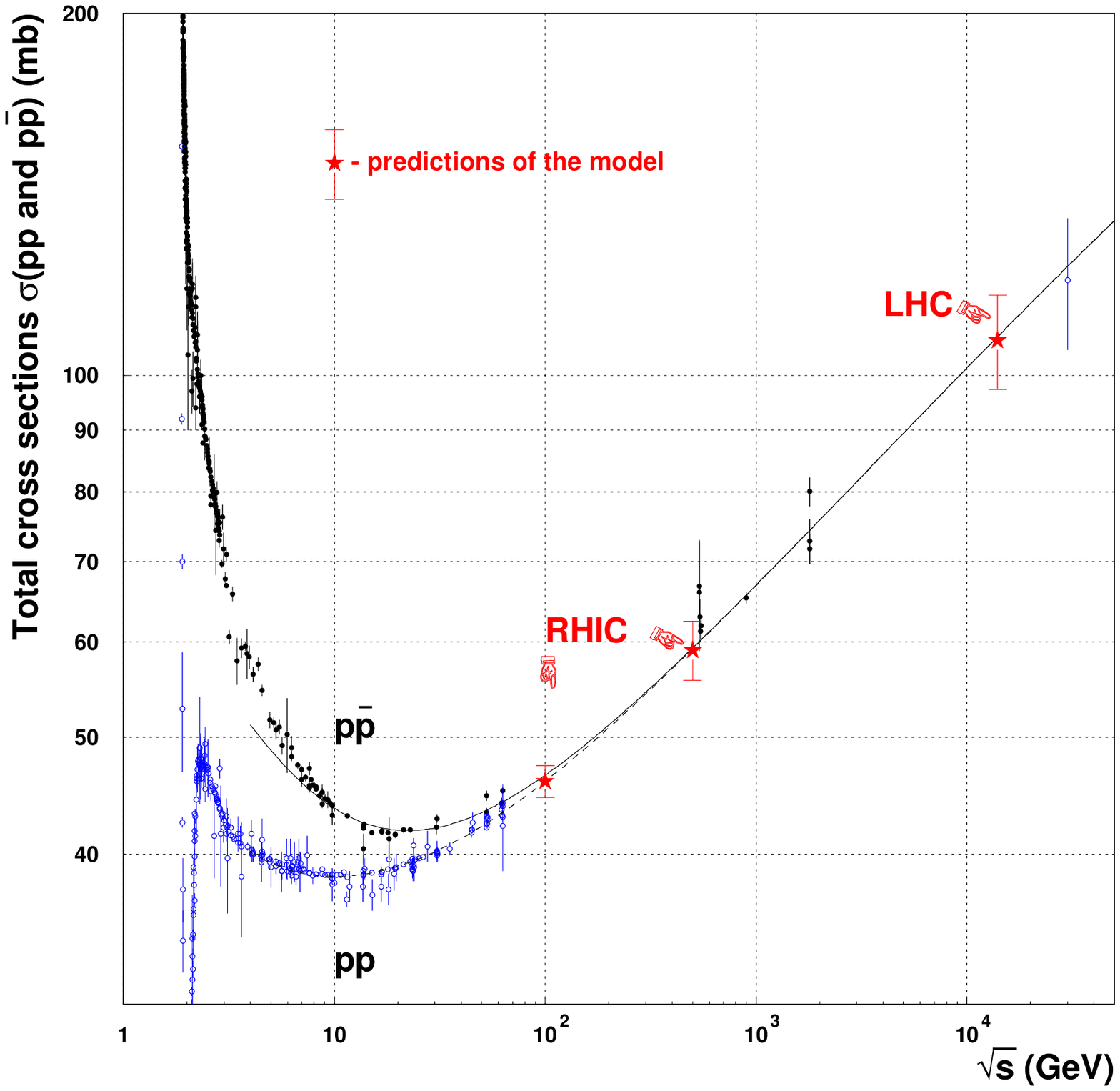}}
\hfill~\parbox[t]{6.cm}{\vspace*{ 0cm} \epsfxsize=70mm \epsffile{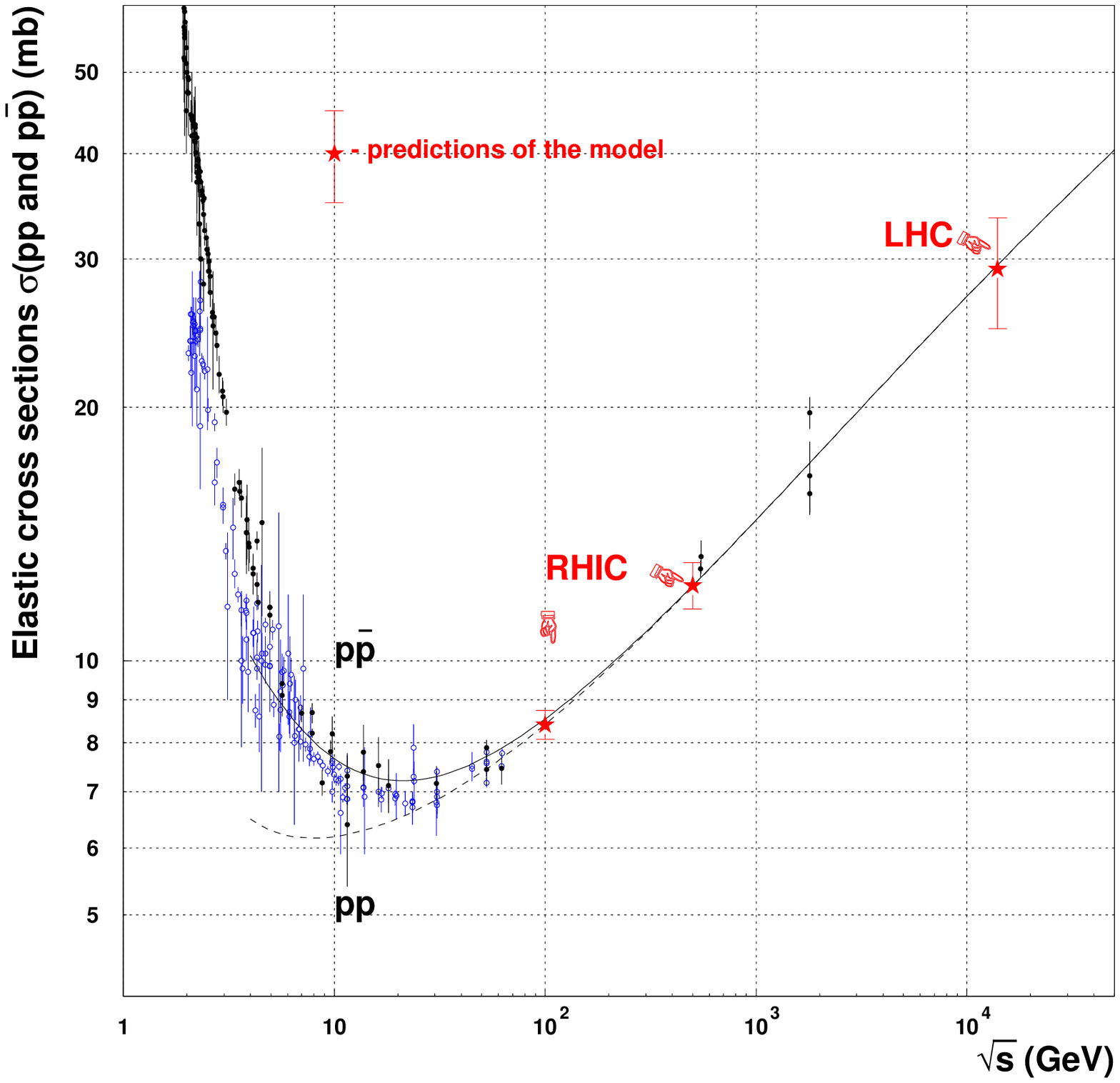}}
\vskip -2.1cm
\parbox[t]{6.5cm}{\caption{Total cross sections of $pp$ scattering 
(hollow circles)  and $\bar p p$ scattering (full circles)
and curves corresponding to their description in the three-Pomeron model.
\label{fig:tot}}}
\hfill~\parbox[t]{6.5cm}{\caption{Elastic cross sections of $pp$ scattering 
(hollow circles)  and $\bar p p$ scattering (full circles)
and curves corresponding to their description in the three-Pomeron model.
These sets of data are not included in the fit.
\label{fig:elastic}
}}
\vskip -1.2cm
\parbox[t]{6.cm}{\vspace*{ 0cm} \epsfxsize=70mm \epsffile{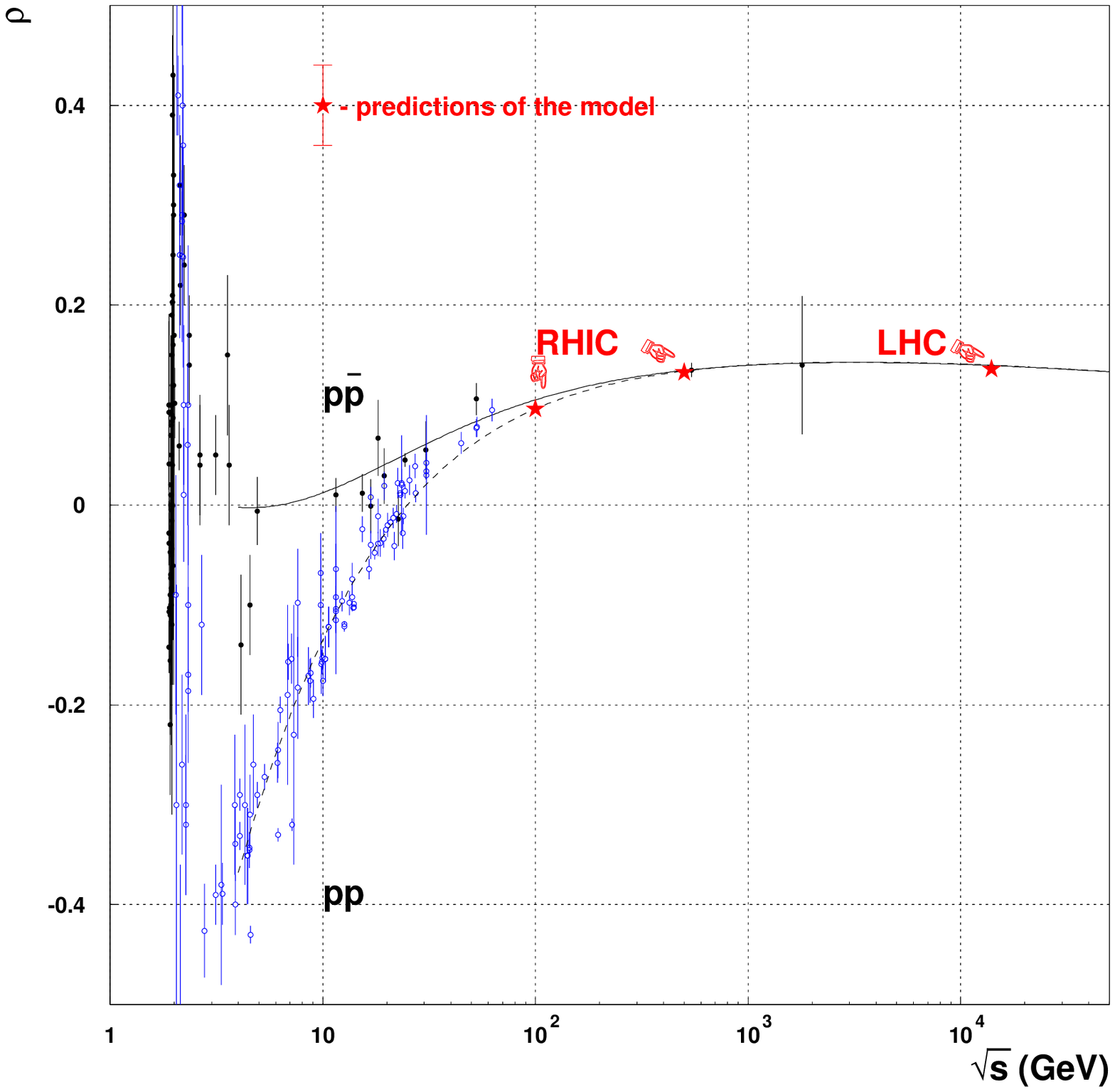}}
\hfill~\parbox[t]{6.cm}{\vspace*{ 0cm} \epsfxsize=70mm \epsffile{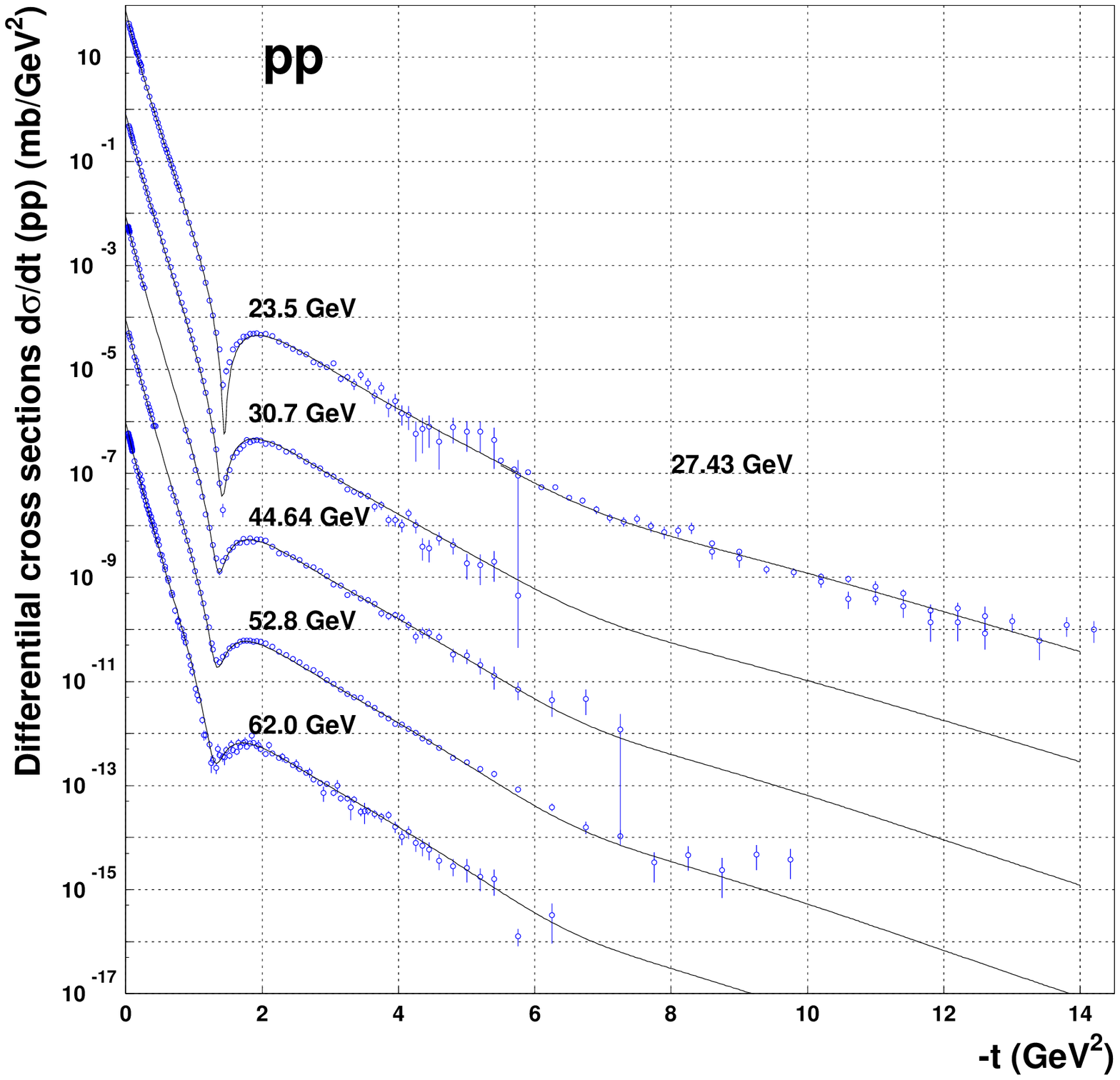}}
\vskip -2.cm
\parbox[t]{6.5cm}{\caption{Ratios of the real to the imaginary part of the forward $pp$ scattering amplitude 
(hollow circles)  and $\bar p p$ scattering  amplitude (full circles)
and curves corresponding to their description in the three-Pomeron model.
\label{fig:reim}
}}
\hfill~\parbox[t]{6.cm}{\caption{Differential cross-sections for $pp$ scattering
and curves corresponding to their description in the three-Pomeron model. 
A $10^{-2}$ factor between each successive set of data is omitted. 
\label{fig:ppdif}
}}
\end{figure}

\begin{figure}[H]
\parbox[t]{6.cm}{\vspace*{ -2cm} \epsfxsize=70mm \epsffile{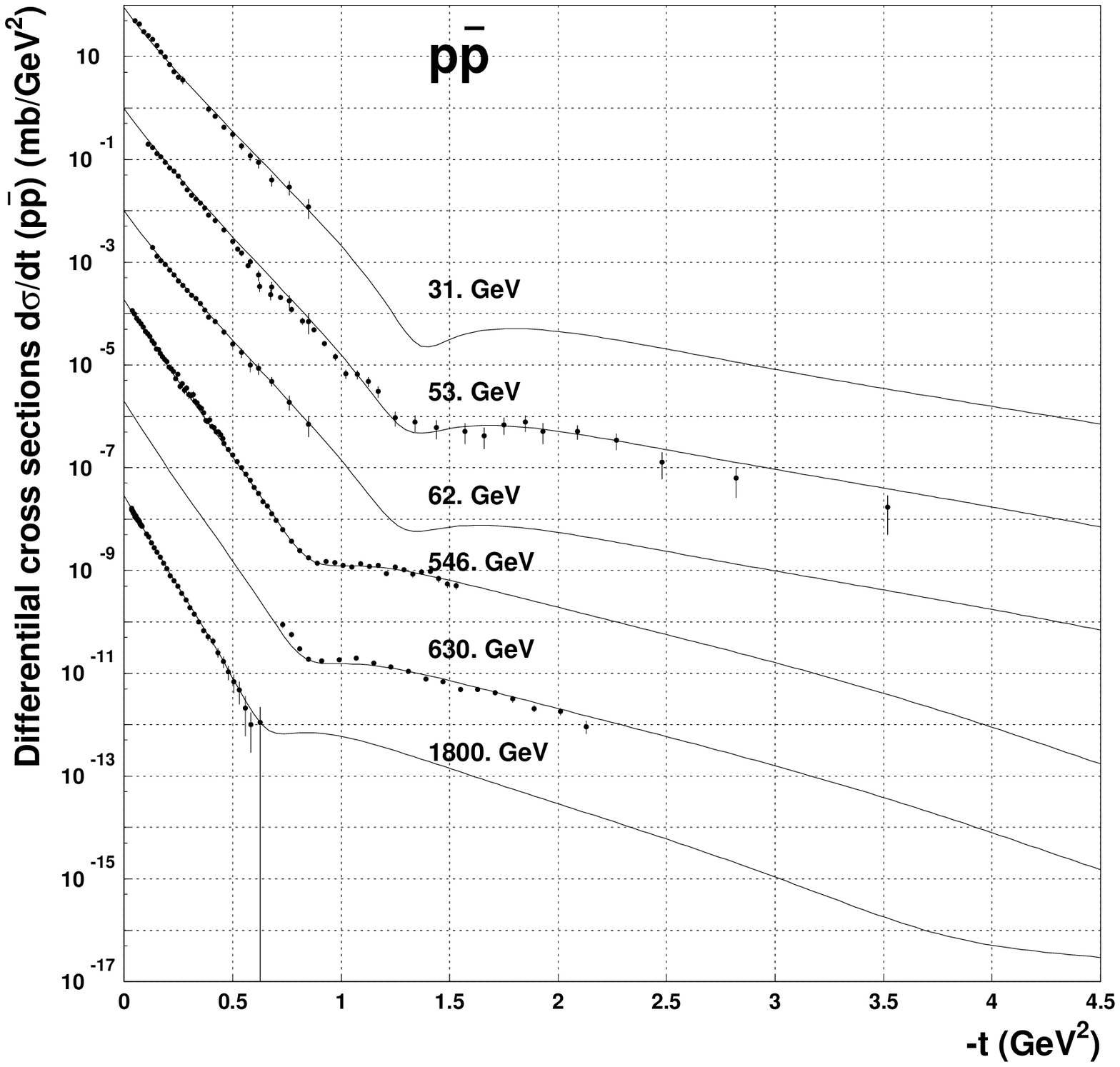}}
\hfill~\parbox[t]{6.5cm}{\vspace*{ -2cm} \epsfxsize=70mm \epsffile{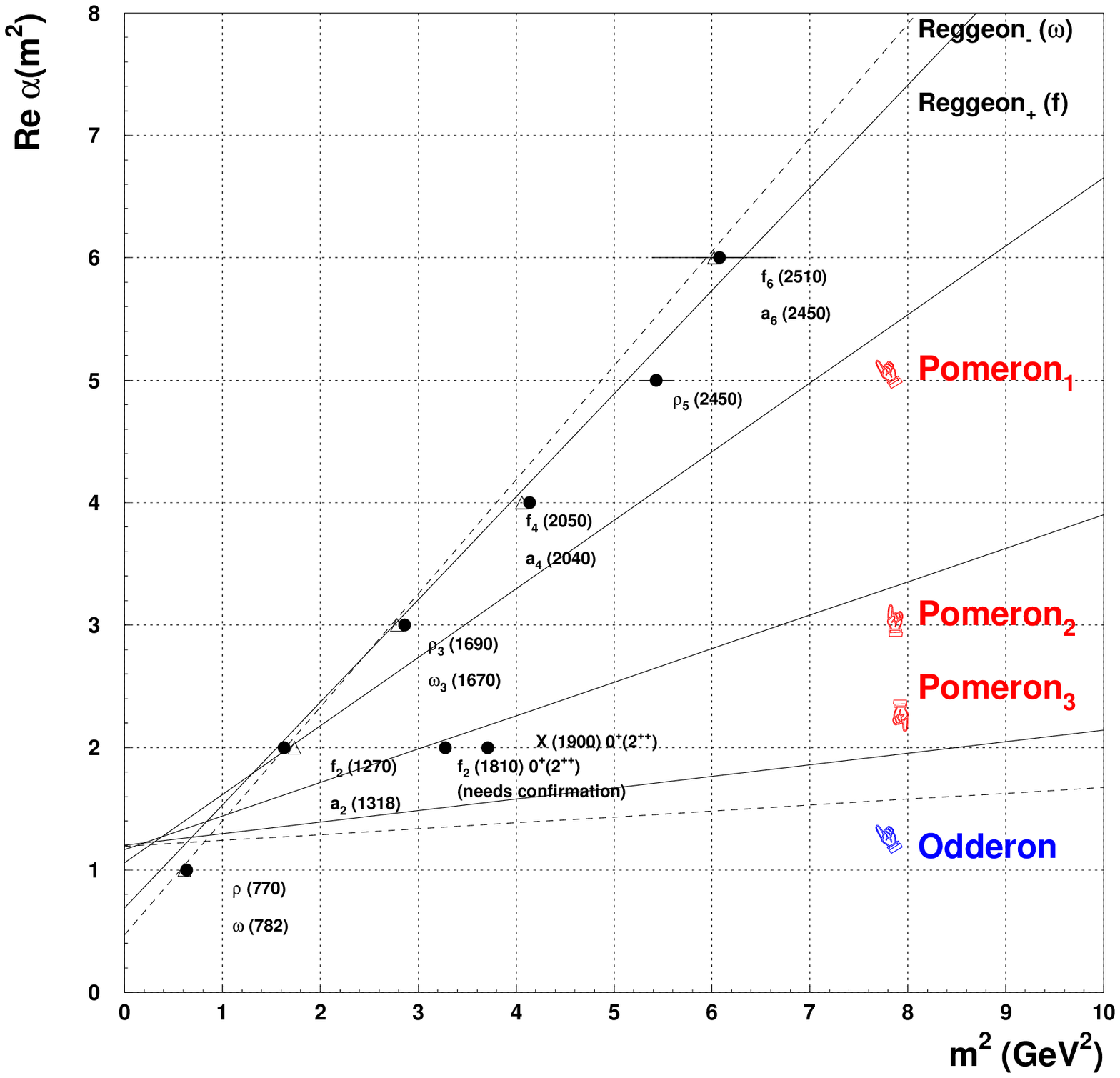}}
\vskip -1.cm
\parbox[t]{6.cm}{\caption{Differential cross-sections for $\bar p p$ scattering
and curves corresponding to their description in the three-Pomeron model. 
A $10^{-2}$ factor between each successive set of data is omitted. 
\label{fig:difpbarp}}}
\hfill~\parbox[t]{6.cm}{\caption{Regge trajectories of secondary Reggeons, three Pomerons and the Odderon.
\label{fig:traject}}
}
\end{figure}

\begin{figure}[H]
\centering
{\vspace*{ -2cm} \epsfxsize=90mm \epsffile{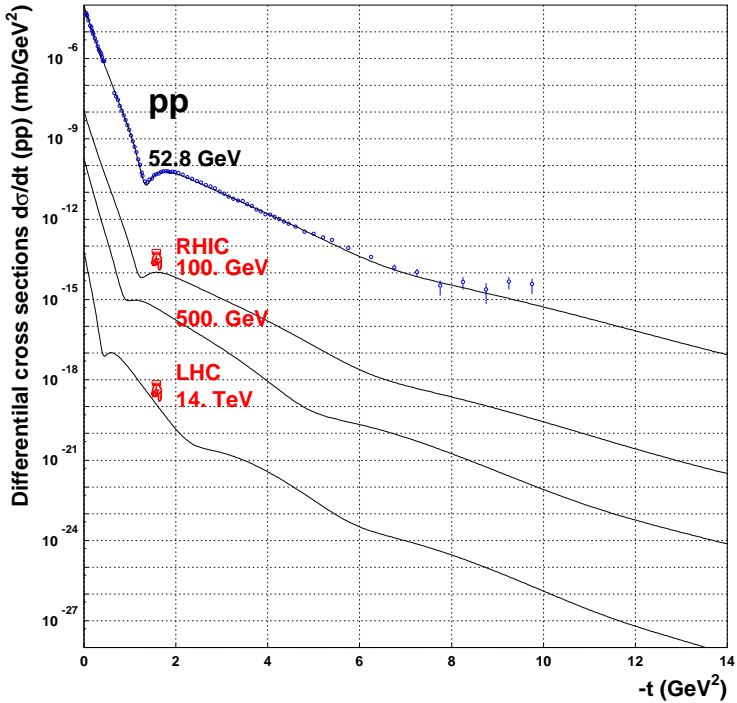}}
\vskip -2.cm
\caption{Predictions of the three-Pomeron model for the differential cross-section of 
$p p$ scattering 
which will be measured at LHC with $\sqrt{s}=14.\; TeV$ and at 
RHIC $\sqrt{s}=100.\; GeV$ and $\sqrt{s}=500.\; GeV$. The data 
corresponding to the energy $\sqrt{s}=52.8\; GeV$ is multiplied by
 $10^{-6}$, RHIC at $500\; GeV$ by $10^{-10}$, RHIC at $500\; GeV$ by $10^{-12}$, and that of LHC by $10^{-16}$.
\label{fig:difpplhc}}
\end{figure}

Having used 20 adjustable parameters we achieved $\chi^2/{\rm d.o.f.}=2.60$. 
(Interested reader may find the list of parameters in the paper \cite{threepomerons}). The results are shown in fig.
\ref{fig:tot}, \ref{fig:elastic}, \ref{fig:reim}, \ref{fig:ppdif},
\ref{fig:difpbarp},
\ref{fig:difpplhc}. We do not include elastic cross-section data sets into the fit and
predictions of the model for elastic
cross-sections can be seen in fig.~\ref{fig:elastic}. 

The model predicts the appearance of two dips in the differential cross-section 
which will be measured at LHC, fig.~\ref{fig:difpplhc}. These dips are
to appear in the region $t_1\simeq -0.5\; GeV^2$ and $t_2\simeq -2.5\; GeV^2$
which is in agreement with other predictions 
(models~\cite{multipomeronPredazzi}, \cite{Wu}).

In  the high $|t|$ domain the model shows predominance of the Odderon contribution and its interference with $Pomeron_3$ contribution, and this predominance of the Odderon is in agreement with a model~\cite{Islam} based on different from ours assumptions.

We predict the following values of the total cross-section, elastic cross-section, and the ratio of 
 real to imaginary part of the amplitude for the LHC, $ \sqrt{s}=14.\; TeV$:\\
$ 
\sigma_{tot}^{pp} = 106.73\;\;(mb)\;_{-\; 8.50mb}^{+7.56 \;mb}\;, 
\sigma_{elastic}^{pp} = 29.19\;\;(mb)\;_{-2.83\; mb}^{+3.58 \;mb}\;, 
\rho^{pp} = 0.1378\;_{-0.0061}^{+0.0042}\;.
$

Predictions for RHIC are:\\
$
\sqrt{s}=100.\; GeV \;:
$ \\
$
\sigma_{tot}^{pp} = 45.96\;\;(mb)\;_{-1.38\; mb}^{+1.41 \;mb}\;, 
\sigma_{elastic}^{pp} = 8.40\;\;(mb)\;_{-0.32\; mb}^{+0.34 \;mb}\;, 
\rho^{pp} = 0.0962\;_{-0.0032}^{+0.0032}\;.
$ \\
$
\sqrt{s}=500.\; GeV \;:
$ \\
$
\sigma_{tot}^{pp} = 59.05\;\;(mb)\;_{-3.10\; mb}^{+2.94 \;mb}\;, 
\sigma_{elastic}^{pp} = 12.29\;\;(mb)\;_{-0.76\; mb}^{+0.79 \;mb}\;, 
\rho^{pp} = 0.1327\;_{-0.0071}^{+0.0052}\;.
$

  The parameters of the Pomeron trajectories are:
\bea
\nonumber
\alpha(0)_{{\Bbb P}_1}=1.058,\;\;\alpha'(0)_{{\Bbb P}_1}=0.560\;(GeV^{-2}); \\
\alpha(0)_{{\Bbb P}_2}=1.167,\;\;\alpha'(0)_{{\Bbb P}_2}=0.273\;(GeV^{-2}); \\
\nonumber
\alpha(0)_{{\Bbb P}_3}=1.203,\;\;\alpha'(0)_{{\Bbb P}_3}=0.094\;(GeV^{-2}). 
\eea

\section{CONCLUSION AND DISSCUSSION}

It is interesting to enlist the following characteristic properties 
of the Pomerons used in this paper.

The first of the Pomerons (`$Pomeron_1$') possesses the properties that we expect
from the string picture~\cite{Soloviev} of Reggeons, 
i.e. $\alpha'(0)_{{\Bbb P}}=\frac{1}{2}\alpha'(0)_{f}=0.42\;(GeV^{-2})$ 
and indeed $\alpha'(0)_{{\Bbb P}_1}=0.559\pm 0.078\;(GeV^{-2})$.

The second Pomeron (`$Pomeron_2$') is close to what is called ``supercritical Pomeron"
with the slope $\alpha'(0)_{{\Bbb P}_2}=0.273\pm0.005\;(GeV^{-2})$ close to its ``world'' value $\alpha'(0)_{{\Bbb P}}\simeq 0.25\;(GeV^{-2})$.
 
The third Pomeron (`$Pomeron_3$') is reminiscent of what is known as a ``hard''
(or perturbative QCD) Pomeron.
Its parameters 
($\alpha(0)_{{\Bbb P}_3}=1.203$, 
$\alpha'(0)_{{\Bbb P}_3}=0.094\; (GeV^{-2})$)
are close to the calculated parameters of the perturbative Pomeron, 
which arise from the summation of reggeized gluon ladders 
and BFKL equation~\cite{bfkl}: $\alpha(0)_{{\Bbb P}}^{BFKL}\simeq 1.2,\;\;\alpha'(0)_{{\Bbb P}}^{BFKL}\sim 0.\;(GeV^{-2})$. 
The fact of arising of a ``hard" Pomeron in a presumably ``soft" framework can seem quite unexpected.
However we are not particularly inclined to identify straightforwardly 
``our hard Pomeron" with that
which is a subject of perturbative QCD studies.

The Odderon has the following parameters: $\alpha(0)_{\Bbb O}=1.192,\;\;\alpha'(0)_{{\Bbb O}}=0.048\;$ $(GeV^{-2})$ in agreement with unitarity constraints 
\cite{pomeronodderon}.
The Odderon intercept is positive and close to that of the $Pomeron_3$. The slope
is almost zero. The coupling is so small that only high-t data may be sensible
to the Odderon contribution.

Assuming that one can neglect the non-linearities of Regge trajectories and making use of a simple
parametrization
\be
\alpha(m^2)=\alpha(0)+\alpha'(0)\cdot m^2\;,
\ee
we can try to estimate the corresponding spectroscopic content of our model.

Then $\Re {\rm e}\alpha(m^2)=J$, where $J$ is an integer number corresponding 
to the spin of a particle which we should find lying on the trajectory. 

The trajectories are depicted in fig.~\ref{fig:traject}. The $C+$ Reggeon trajectory is in fact a combination of two families of mesons $f$ and $a_2$. 
The $C-$ Reggeon trajectory is a combination of two families of mesons $\omega$ and $\rho$. As is seen, the secondary Reggeon trajectories fairly well describe the spectrum of mesons.

Among the mesons with appropriate quantum numbers there exesit two that fit
the Pomeron trajectory ($0^+J^{++}$): $f_2(1810)\;\;0^+2^{++}$ with mass $m=1815\pm12 \;\; MeV$ and $X(1900)\;\;0^+2^{++}$ with mass $m=1926\pm12 \;\; MeV$
. One of them is supposed to be on $Pomeron_2$ trajectory.

Returning to our ``polypomeron" hypothesis it is fairly natural to ask what happens if one admits
fourth etc Pomeron? Is there some optimum in the number of ``relevant" Pomerons above which 
the quality of description is not improved much? And, finally, what is underlying physics of 
such a construction?
We hope very much to be able to answer at least some of these questions in the nearest future.

\section*{ACKNOWLEDGMENTS}
One of us A. P. is grateful to the Organizing Committee and to Vojt\v{e}ch Kundr\'{a}t personally for the kind invitation and perfectly organized conference. 
We would like to thank Maurice Giffon for comments, Enrico Predazzi for useful discussions
and for critical reading of the manuscript and Munir Islam for friutful disscussions and suggestions improving the presentation.


\end{document}